\documentclass[pra,twocolumn,showpacs]{revtex4}
\usepackage{amsmath}
\usepackage{graphicx}
\usepackage[usenames]{color}
\usepackage{amssymb}

\begin{document}

\title{Competition between symmetry breaking and onset of collapse \\ in
weakly coupled atomic condensates}
\author{L. Salasnich$^{1}$, B.A. Malomed$^{2}$ and F. Toigo$^{1}$}
\affiliation{$^1$CNISM and CNR-INFM, Unit\`a di Padova, Dipartimento di 
Fisica ``G.
Galilei'', Universit\`a di Padova, Via Marzolo 8, 35131 Padova, Italy \\
$^{2}$Department of Physical Electronics, School of Electrical Engineering,
Faculty of Engineering, Tel Aviv University, Tel Aviv 69978, Israel}

\begin{abstract}
We analyze the symmetry breaking of matter-wave solitons in a pair of
cigar-shaped traps coupled by tunneling of atoms. The model is based on a
system of linearly coupled nonpolynomial Schr\"{o}dinger equations (NPSEs).
Unlike the well-known spontaneous-symmetry-breaking (SSB) bifurcation in
coupled cubic equations, in the present model the SSB competes with the
onset of collapse in this system. Stability regions of symmetric and
asymmetric solitons, as well as the collapse region, are identified in the
parameter space of the system.
\end{abstract}

\maketitle

\textit{Introduction}. It is commonly known that the Gross-Pitaevskii
equation (GPE) furnishes a very accurate description of the Bose-Einstein
condensation (BEC) in rarefied gases of bosonic atoms \cite{GPE}. One 
important applications of the GPE is the prediction of Josephson
oscillations \cite{JJ} and \textit{spontaneous symmetry breaking} (SSB),
alias macroscopic quantum self-trapping \cite{bifurcation}, in double-well
potentials (DWPs). Experimentally, the oscillations \cite%
{experiment,Heidelberg} and self-trapping \cite{Heidelberg} have been
demonstrated in BECs with repulsive interactions between atoms, see Ref.
\cite{review} for a review. Asymmetric stationary states trapped in the DWP
are generated by symmetry-breaking bifurcations from symmetric or
antisymmetric states, in case the intrinsic nonlinearity is attractive or
repulsive, respectively \cite{bifurcation}. In fact, this bifurcation was
first predicted (for the self-focusing nonlinearity) in the model of
dual-core nonlinear optical fibers \cite{dual-core}.

A natural extension of the DWP is a double-channel structure in the
two-dimensional (2D) geometry, based on two potential wells in one direction
$(x)$, which are uniformly extended, as parallel troughs, in the
perpendicular direction $(z)$ \cite{Warsaw,Arik}. With the attractive cubic
nonlinearity, this structure gives rise to double-hump solitons, which are
supported by the DWP in the direction of $x$, while being self-trapped along
$z$, as the ordinary matter-wave solitons in the cigar-shaped (single-core)
trap \cite{solitons}. Similarly shaped gap solitons were predicted in the
double-channel model with the self-repulsive nonlinearity and a periodic
potential (an optical lattice, OL) acting along $z$ \cite{Arik,Warsaw2}. If
the nonlinearity is strong enough, an obvious symmetric/antisymmetric
soliton in the double-channel setting with the self-attraction/repulsion may
bifurcate into an asymmetric mode. This was demonstrated both in the full 2D
models \cite{Warsaw,Warsaw2} and their 1D counterparts, which replace the
single 2D GPE by a pair of 1D equations with coordinate $x$, for wave
functions in the two potential troughs, while the tunneling between the
troughs in the $x$ direction is approximated by a linear coupling between
the 1D GPEs \cite{Arik}. For the attractive nonlinearity, the system of 1D
GPEs is identical to the model of dual-core nonlinear optical fibers with
anomalous group-velocity dispersion, where the symmetry-breaking bifurcation
and asymmetric solitons generated by it have been studied in detail \cite%
{Sydney,Akhmed}. The difference between the linearly-coupled models with the
intrinsic attraction and repulsion is in the character of the SSB\
bifurcation, which is \textit{subcritical} and \textit{supercritical }in the
former and latter cases, respectively. The subcritical bifurcation gives
rise to a narrow region of the bistability, where stable symmetric and
asymmetric solitons coexist \cite{Sydney}.

The DWP may also be extended in two transverse directions, forming a pair of
parallel pancakes-shaped traps. The model developed for this setting is
based on a pair of linearly-coupled 2D GPEs, which give rise to the SSB of
2D solitons and solitary vortices, for either sign of the nonlinearity,
provided that the model includes an in-plane OL potential (otherwise, all 2D
solitons are unstable) \cite{Arik2}.

The analyses of the SSB in solitons, reported in previous works, were
dealing with the cubic nonlinearity, except for Ref. \cite{Albuch}, which
considered a system of coupled equations combining the cubic attractive and
quintic repulsive terms -- a setting relevant to optics, rather than to BEC.
It was demonstrated that the bifurcation diagrams, accounting for the SSB of
solitons in that system, form a closed loop, unlike open diagrams generated
by the cubic nonlinearity. An essential aspect of the description of BEC in
the low-dimensional setting is that the reduction of the GPE from 3D to 1D
transforms the original cubic nonlinearity into a \emph{nonpolynomial} form
\cite{Luca,Canary}. In the case of the self-attraction, the respective
\textit{nonpolynomial nonlinear Schr\"{o}dinger} equation (NPSE) predicts
the occurrence of collapse at a critical value of the density, in agreement
with the underlying 3D cubic GPE \cite{Luca}. The objective of this work is
to study a hitherto unexplored aspect of the SSB in solitons, \textit{viz}.,
the competition between the SSB and the onset of the collapse, in the
framework of a system of linearly coupled NPSEs (in Refs. \cite{Arik2}, the
possibility of the collapse of asymmetric solitons in the 2D dual-core model
with the attractive cubic nonlinearity and OL potential was mentioned, but
not investigated, as the SSB happened there at much lower values of the
density than those necessary for the onset of the collapse). Below, we
derive the NPSE system, and then produce a diagram in its parameter space,
which reveals regions of stable symmetric and asymmetric solitons and a
collapse area.

\textit{The model}. The starting point of the analysis is the scaled 3D GPE
for the mean-field wave function, $\psi $, which describes the BEC in two
parallel identical cigar-shaped traps separated by a potential barrier:%
\begin{eqnarray}
i\psi _{t} &=&\frac{1}{2}\left\{ -\nabla ^{2}+\Omega ^{2}\left[ \left(
x-a\right) ^{2}+\left( x+a\right) ^{2}-4a^{2}+y^{2}\right] \right\} \psi
\notag \\
&+&W(z)\psi -2\pi g\left\vert \psi \right\vert ^{2}\psi ,  \label{GPEs}
\end{eqnarray}%
where $z$ is the longitudinal coordinate, $W(z)$ is the axial
potential (if any), $y$ is directed perpendicular to the plane drawn
through axes of the parallel traps, $x$ lies in the plane, being
orientated perpendicular to the axes, and $2a$ is the distance
between them. The coefficient $2\pi g>0$ accounts for the attraction
between atoms, and $\Omega ^{2}$ -- for the transverse isotropic
trapping in each ``cigar". 

Our first objective is to reduce Eq. (\ref{GPEs}) to a system of linearly
coupled 1D equations. To this end, we modify the approach developed for the
single ``cigar" \cite{Luca}, adopting a superposition of two single-core
\textit{ans\"{a}tze}:%
\begin{eqnarray}
\psi (x,y,z,t) &=&\frac{1}{\sqrt{\pi }}\left[ \exp \left( -\frac{\left(
x-a\right) ^{2}+y^{2}}{2\sigma _{1}^{2}}\right) \frac{f_{1}(z,t)}{\sigma _{1}%
}\right.  \notag \\
&&\left. +\exp \left( -\frac{\left( x+a\right) ^{2}+y^{2}}{2\sigma _{2}^{2}}%
\right) \frac{f_{2}(z,t)}{\sigma _{2}}\right] .  \label{ansatz}
\end{eqnarray}%
Here $f_{1}$ and $f_{2}$ are the 1D (axial) wave functions in the two cores,
and $\sigma _{1,2}$ are the respective transverse widths.

We proceed by substituting ansatz (\ref{ansatz}) into the energy functional
(Hamiltonian) corresponding to Eq. (\ref{GPEs}),
\begin{gather}
E=\frac{1}{2}\int \int \int dxdydz\left\{ \left\vert \nabla \psi \right\vert
^{2}+\Omega ^{2}\left[ \left( x-a\right) ^{2}+\right. \right.  \notag \\
\left. \left. \left( x+a\right) ^{2}-4a^{2}+y^{2}\right] |\psi
|^{2}+2W(z)\left\vert \psi \right\vert ^{2}-2\pi g|\psi |^{4}\right\} .
\label{E}
\end{gather}%
The underlying assumption, that the distance between the ``cigars" is
essentially larger than the radius of the transverse confinement in each of
them, implies $a^{2}\Omega \gg 1.$ Due to this condition, the coupling
energy, which is produced by the overlap of the two components of the wave
function in ansatz (\ref{ansatz}), if substituted into Hamiltonian (\ref{E}%
), takes the following form, for $\sigma _{1}=\sigma _{2}\equiv \sigma $: $%
E_{\mathrm{coupl}}=-\kappa \int_{-\infty }^{+\infty }\left[
f_{1}(z)f_{2}^{\ast }(z)+f_{1}^{\ast }(z)f_{2}(z)\right] dz,$ where $\kappa
\equiv 2\left( a\Omega \right) ^{2}\exp \left( -a^{2}/\sigma ^{2}\right) .$%
The main contribution to $E_{\mathrm{coupl}}$ comes from region $%
x^{2},y^{2}\lesssim \sigma ^{2}$ around the midpoint between the ``cigars".
In that region, the transverse- confinement radius is determined by the
ground-state wave function of the 2D harmonic oscillator, i.e., $\sigma
=\Omega ^{-1/2}$, hence the coupling coefficient becomes a constant, $\kappa
=2\left( a\Omega \right) ^{2}\exp \left( -a^{2}\Omega \right) .$

Other terms in Hamiltonian (\ref{E})\ are calculated separately for $f_{1}$
and $f_{2}$. The eventual result is (from now on, we set $\Omega \equiv 1$
by means of an obvious rescaling)
\begin{eqnarray}
E &=&(1/2)\sum_{n=1,2}\int_{-\infty }^{+\infty }\left\{ \left( \sigma
_{n}^{-2}+\sigma _{n}^{2}+2W(z)\right) \left\vert f_{n}(z)\right\vert
^{2}\right.  \notag \\
&&\left. +\left\vert \partial f_{n}/\partial z\right\vert ^{2}-g\sigma
_{n}^{-2}\left\vert f_{n}(z)\right\vert ^{4}\right\} dz  \notag \\
&-&\kappa \int_{-\infty }^{+\infty }\left[ f_{1}(z)f_{2}^{\ast
}(z)+f_{1}^{\ast }(z)f_{2}(z)\right] dz\;,  \label{Efinal}
\end{eqnarray}%
where, as usual \cite{Luca}, the $z$-dependence of $\sigma _{n}$ is
neglected. The transverse widths are determined by the variational
equations, $\delta E/\delta \sigma _{n}=0$, which leads to the same
relations as in the single-core model: $\sigma _{n}^{2}=\sqrt{1-g\left\vert
f_{n}\right\vert ^{2}}.$ The substitution of this into energy functional (%
\ref{Efinal}) casts it into the final form,

\begin{gather}
E=\int_{-\infty }^{+\infty }\left\{ \left[ \sum_{n=1,2}\left( \sqrt{%
1-g\left\vert f_{n}\right\vert ^{2}}+W(z)\right) \left\vert
f_{n}(z)\right\vert ^{2}\right. \right.   \notag \\
\left. \left. +\frac{1}{2}\left\vert \frac{\partial f_{n}}{\partial z}%
\right\vert ^{2}\right] -\kappa \left[ f_{1}(z)f_{2}^{\ast }(z)+f_{1}^{\ast
}(z)f_{2}(z)\right] \right\} dz.  \label{simple}
\end{gather}%
The system of coupled NPSEs is derived from Hamiltonian (\ref{simple}) as $%
i\left( f_{n}\right) _{t}=\delta E/\delta f_{n}^{\ast }$ $,$ i.e.,
\begin{equation}
i\frac{\partial f_{n}}{\partial t}=-\frac{1}{2}\frac{\partial ^{2}f_{n}}{%
\partial z^{2}}+W(z)f_{n}+\frac{1-\left( 3/2\right) g\left\vert
f_{n}\right\vert ^{2}}{\sqrt{1-g\left\vert f_{n}\right\vert ^{2}}}%
f_{n}-\kappa f_{3-n},  \label{f}
\end{equation}%
Note that the derivation of the equations for $f_{n}$ from the Hamiltonian
in the form of Eq. (\ref{Efinal}), and subsequent substitution of
expressions $\sigma _{n}^{2}=\sqrt{1-g\left\vert f_{n}\right\vert ^{2}\text{,%
}}$ leads to the same equations (\ref{f}). Equations (\ref{f}) conserve
energy (\ref{simple}) and the total norm (number of atoms in the
condensate), $N=\int_{-\infty }^{+\infty }\left( \left\vert
f_{1}(z)\right\vert ^{2}+\left\vert f_{2}(z)\right\vert ^{2}\right) dz\equiv
N_{1}+N_{2}$.

In the low-density limit, $g|f_{1,2}|^{2}\ll 1$, Eqs. (\ref{f}) reduce to 1D
GPEs,
\begin{equation}
i\left( f_{n}\right) _{t}=-(1/2)\left( f_{n}\right)
_{zz}+W(z)f_{n}-g|f_{n}|^{2}f_{n}-\kappa f_{3-n}~,  \label{f2}
\end{equation}%
This system, with $g>0$, and $W(z)=0$, is tantamount to the model of the
dual-core nonlinear optical fiber with the anomalous group-velocity
dispersion, where the SSB of solitons with $f_{1}\left( z,t\right)
=f_{2}\left( z,t\right) $ was studied in detail \cite{Sydney,Akhmed}. In
terms of the double-trap BEC model based on system (\ref{f2}) with the OL
potential, $W(z)=\epsilon \cos (2kx)$, the SSB\ of regular solitons, for $%
g>0 $, and gap solitons, for $g<0$, was studied in Ref. \cite{Arik}. In the
case of $g<0$, the bifurcation actually breaks the \textit{antisymmetry} of
the two-component gap solitons with $f_{1}\left( z,t\right) =-f_{2}\left(
z,t\right) $. In the present work, we focus on the model with the
self-attraction, $g>0$, when the difference of the nonlinearity in Eqs. (\ref%
{f2}) from the cubic form is important. As said above, the most interesting
issue is the competition between the SSB and the onset of the collapse,
which is admitted by the NPSE even in the framework of the 1D description
\cite{Luca}.

\textit{Results}. To specify the normalization for functions $f_{1,2}(z)$, we
fix the total norm, $N_{1}+N_{2}\equiv 2,$ while $N_{1}$ and $N_{2}$ may
vary. Then, the full set of stationary solutions to Eqs. (\ref{f}), $%
f_{1,2}(z,t)=e^{-i\mu t}U_{1,2}(x)$, with real functions $U_{1,2}(z)$ and
chemical potential $\mu $, is parameterized by two coefficients, $g$ and $%
\kappa $. We integrated Eqs. (\ref{f}) in imaginary time, using a
finite-difference Crank-Nicolson code and imposing condition $N=2$ at each
step. Initial wave functions were taken as Gaussians with norms $N_{1}=1.01$
and $N_{2}=N-N_{1}=0.99$.

\begin{figure}[tbp]
\center\includegraphics [width=9.cm,clip]{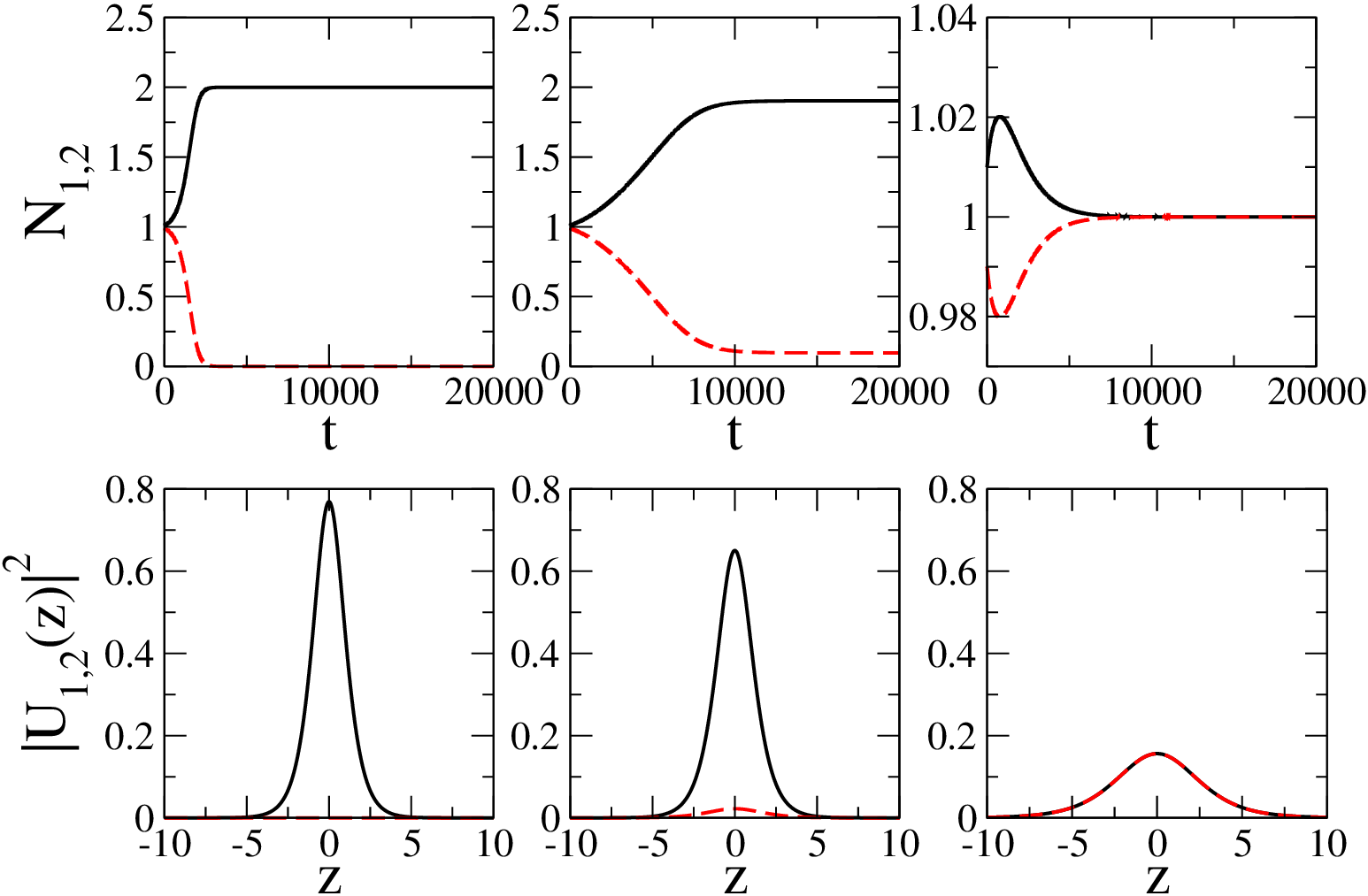}
\caption{(Color online). Upper panels: norms $N_{1}$ and $N_{2}$ (solid and
dashed lines) in the course of the evolution in imaginary time. Lower
panels: final density profiles of $U_{1}(z)$ and $U_{2}(z)$ (solid and
dashed lines), corresponding to the cases shown in the upper panels. The
nonlinearity coefficient is $g=0.6$, while the coupling constant is $\protect%
\kappa =0$, $0.06$, and $0.12$, in the left, central, and right panels,
respectively. }
\label{fig1}
\end{figure}

In Fig. \ref{fig1} we report typical examples of totally asymmetric,
strongly asymmetric, and symmetric solitons found below the collapse
threshold (at $g=0.6$), for different values of the linear coupling $\kappa $. 
At $\kappa =0$,
wave function $f_{1}(z)$ with initial norm $N_{1}=1.01$ evolves into a
single-component bright soliton which absorbs the total norm, $N=2$, while $%
f_{2}(z)$, with initial norm $N_{2}=0.99$, decays to zero. At $\kappa =0.06$
(weak linear coupling), the wave functions evolve into a stable
two-component soliton with a strongly broken symmetry. Finally, the strong
linear coupling, with $\kappa =0.12$, enforces the evolution of the soliton
into the symmetric form.

\begin{figure}[tbp]
\center\includegraphics [width=8.cm,clip]{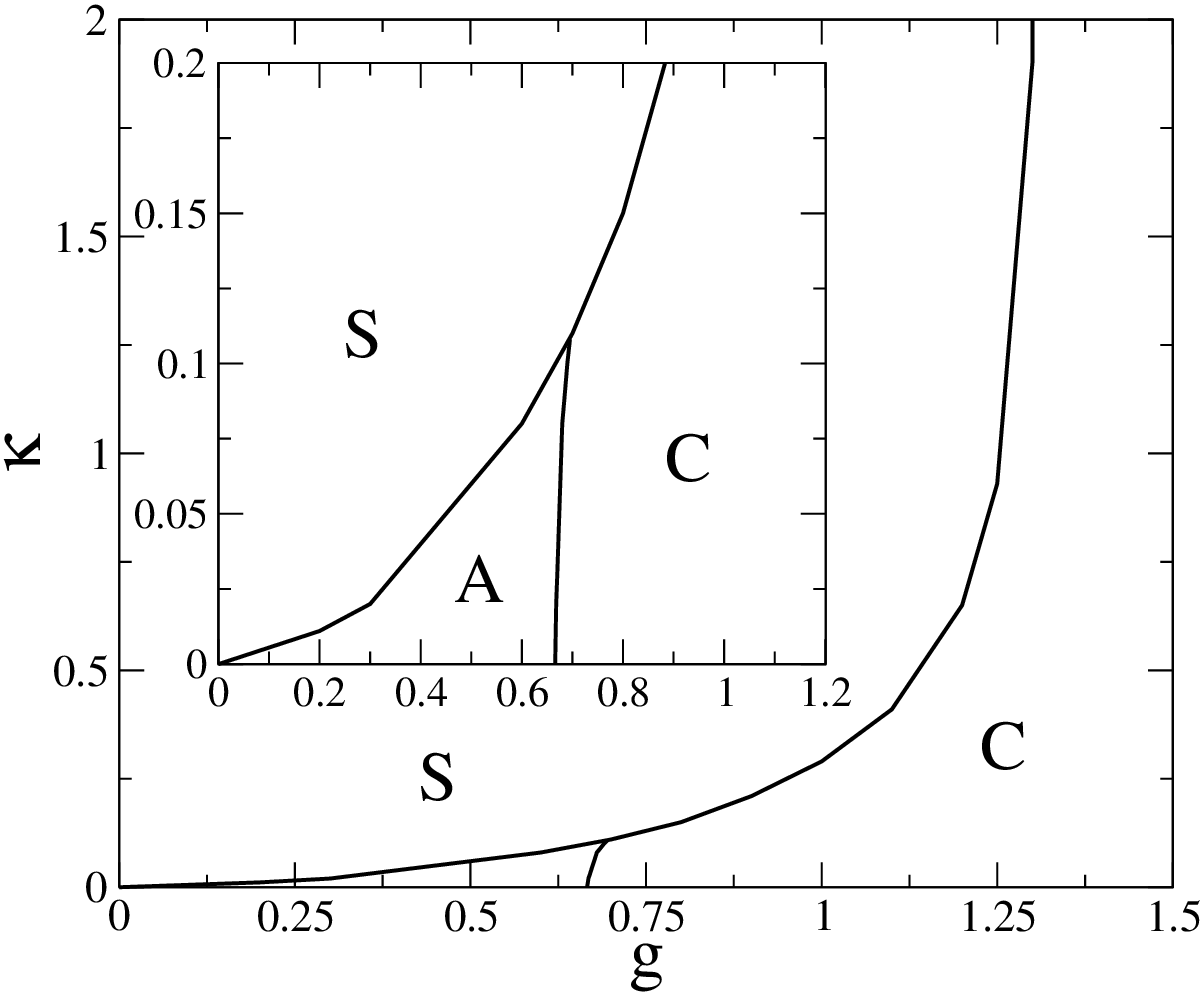}
\caption{The phase diagram of the linearly-coupled system of NPSEs in the
parameter plane. In regions S and A, the system supports, respectively,
stable symmetric [$U_{1}(z)=U_{2}(z)$] and asymmetric [$U_{1}(z)\neq
U_{2}(z) $] stationary solutions. The collapse takes place in region C. The
inset is a zoom of the area at small values of coupling constant $\protect%
\kappa $, which shows the border between A and C in detail.}
\label{fig2}
\end{figure}

The collapse of wave function $f_{1}\left( z,t\right) $ occurs in
simulations of system (\ref{f}) with $\kappa =0$ at $g\geq 2/3$, if the
evolution commences with initial norm $N_{1}=1.01$. This result agrees with
known properties of the single-component NPSE\ \cite{Luca}, assuming that
the total norm, $N=2$, goes to $f_{1}$. A suppression of the collapse takes
place at finite values of $\kappa $. The results are summarized in Fig. \ref%
{fig2}, in the form of a diagram in the plane of ($g$,$\kappa $). It
features three regions: stable symmetric and asymmetric solitons in S and A,
respectively, and collapsing solutions in C. An obvious rescaling argument 
explains that all solutions suffer the collapse at any $\kappa $ for $g\geq
4/3$, which is twice the above-mentioned critical value, $g=2/3$, for the 
single-component system with $N=2$. It corresponds to the symmetric collapse
mode, with equal norms $N_{1}=N_{2}=1$ in each trap. In accordance with 
these features, the border between regions A and C in Fig. \ref{fig2} 
starts, at $\kappa =0$, with $g=2/3$, while the border \ between S and C
asymptotes to $g=4/3$ at large values of $\kappa $. 
Notice that in the inset of Fig. \ref{fig2} the (A,C) border 
seems straight but it is actually curved, as shown in the main figure. 

\begin{figure}[tbp]
\center\includegraphics [width=7.cm,clip]{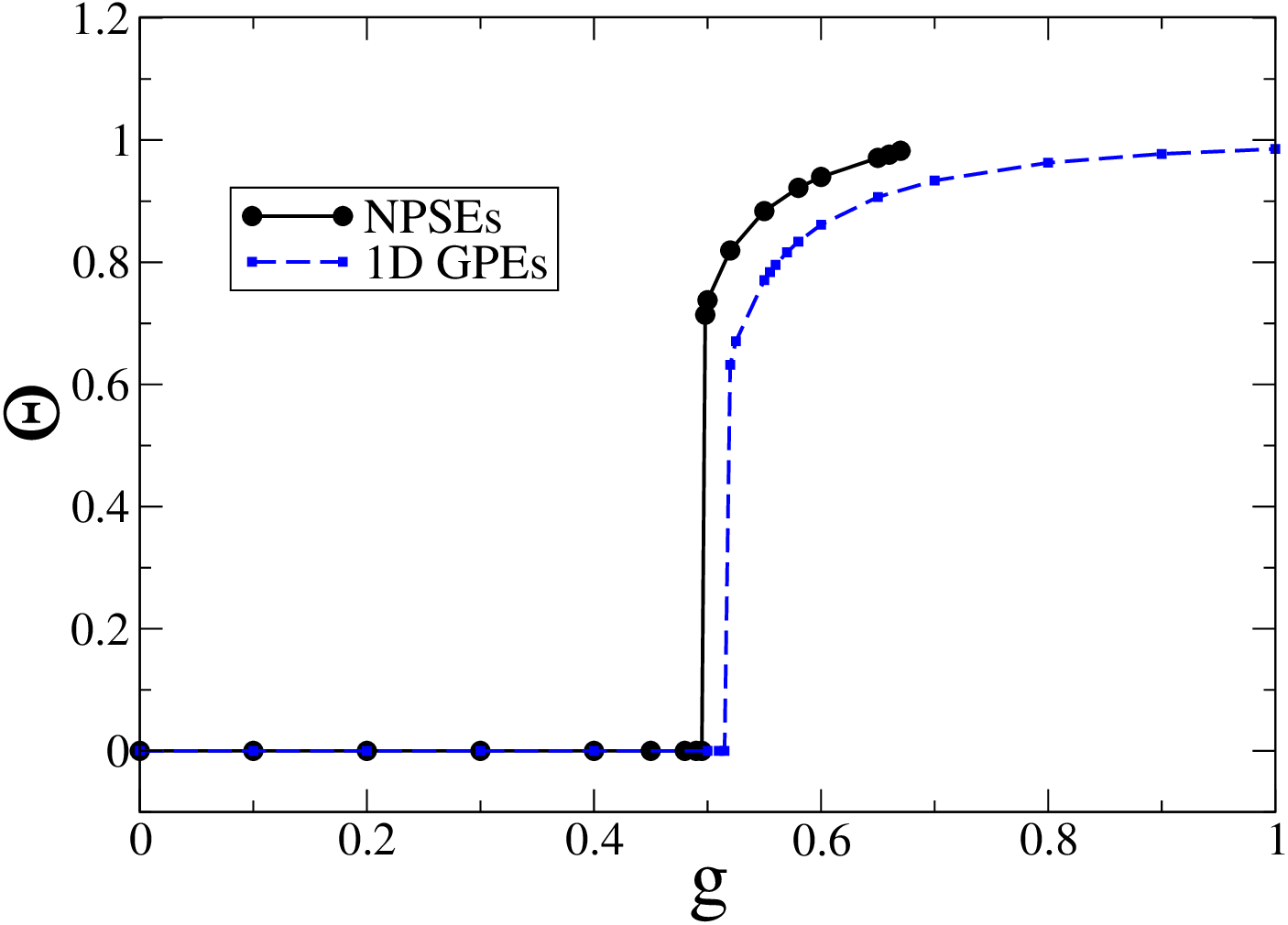}
\center\includegraphics [width=7.cm,clip]{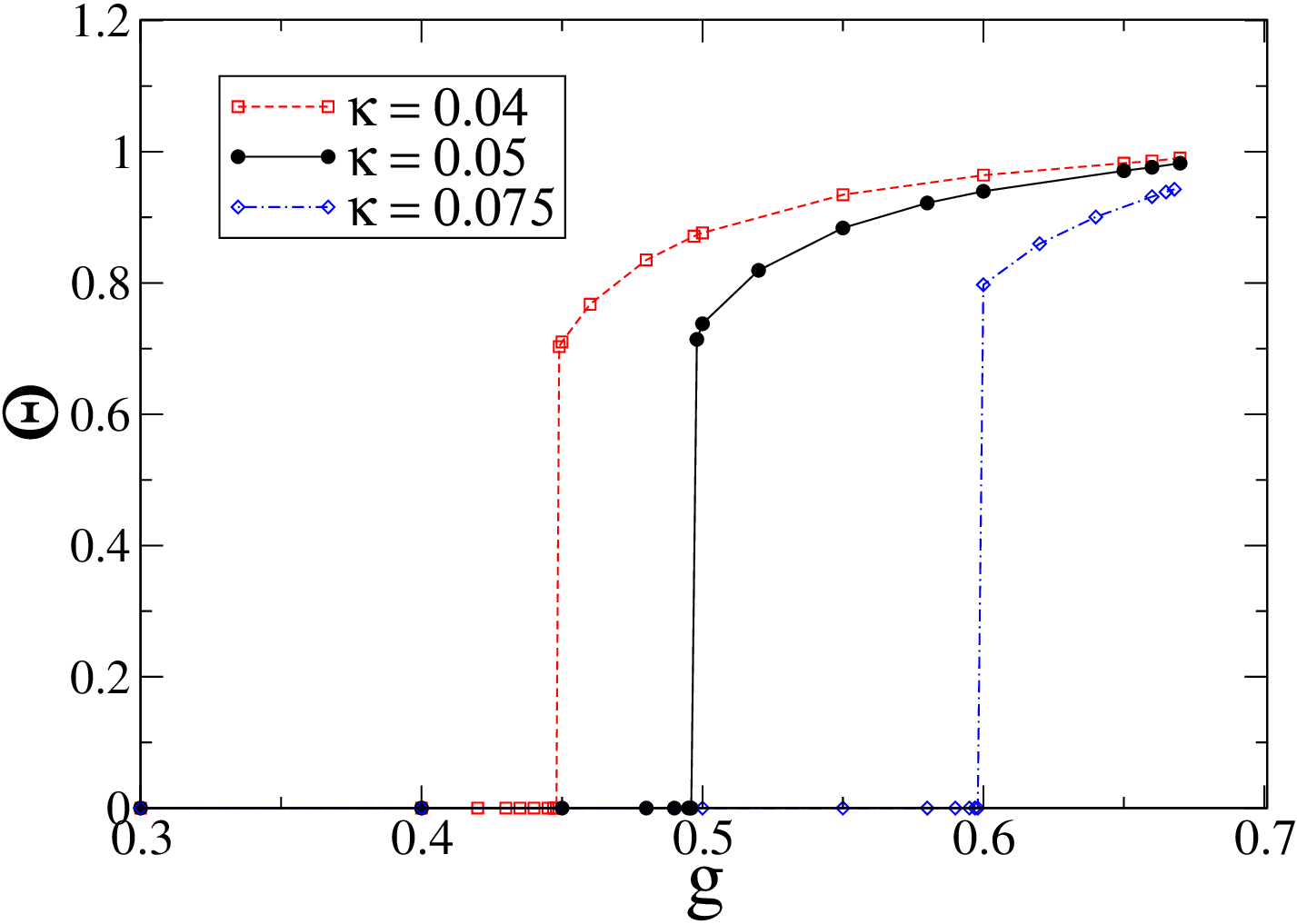}
\caption{(Color online). The asymmetry parameter, $\Theta =\left(
N_{1}-N_{2}\right) /\left( N_{1}+N_{2}\right) $, as a function of
interaction strength $g$. Upper panel: fixed $\protect\kappa =0.05$, 
as obtained from the coupled NPSEs (squares) and 1D GPEs, 
i.e., usual equations with the cubic nonlinearity. 
Lower panel: three values of the 
parameter $\kappa$; curves obtained with the coupled NPSEs. 
The lines corresponding to the NPSE system terminate 
at the collapse point.}
\label{fig3}
\end{figure}

The SSB in solitons is characterized by the asymmetry parameter \cite%
{Sydney,Arik}, $\Theta =\left( N_{1}-N_{2}\right) /\left( N_{1}+N_{2}\right)
.$ The competition between the symmetry breaking and collapse is further
illustrated in Fig. \ref{fig3} by plots of $\Theta $ versus $g$ for a
relatively weak coupling. In the upper panel we fix 
$\kappa =0.05$ and compare the usual model based on linearly coupled 
cubic GPEs, Eqs. (\ref{f2}), with the linearly coupled NPSEs, Eqs. (\ref{f}). 
In the lower panel we plot the NPSE curves for three values of $\kappa$. 

The diagrams of Fig. \ref{fig3} feature a leap from the 
symmetric configuration with $\Theta =0$ to the asymmetric one with $\Theta
\neq 0$, cf. a similar situation in the 2D model considered in Ref. 
\cite{Warsaw}. The transition to asymmetric states in the present model always
happens by a leap, i.e., the symmetry-breaking bifurcation is always 
subcritical, similar to the situation in the coupled equations with the 
self-attractive cubic nonlinearity \cite{Sydney}. 
The SSB may be considered as a first-order 
quantum (zero-temperature) phase transition, with $\Theta $ playing 
the role of the order parameter of the transition \cite{huang}. 
The main difference between the 
$\Theta \left( g\right) $ plots generated by Eqs. (\ref{f}) and Eqs. (\ref{f2}%
) is the fact that the former system predicts the collapse of the asymmetric
configuration at $g\approx 0.66$, while the cubic nonlinearity in the latter
system does not give rise to any collapse. 
Lastly, it is seen from Fig. \ref{fig2} that asymmetric solitons do not
exist at $\kappa >\kappa _{\max }\approx 0.108$. Accordingly, with these
values of the linear-coupling coefficient the SSB does not occur, and the
collapse happens directly in the symmetric mode, at the above-mentioned
critical value, $g=4/3$. 

\textit{Conclusion}. The objective of this work was to explore the SSB in
solitons described by the system of linearly coupled NPSEs. The difference
from the soliton bifurcations in coupled cubic equations is that the SSB
competes with the onset of the collapse. The phase diagram in the system's
parameter plane was produced, revealing regions of stable symmetric and
asymmetric solitons, and of the collapse as well. The symmetry-breaking
bifurcation is subcritical, unless the collapse occurs before the SSB. It
may be quite interesting to extend this model for the description of the SSB
in a pair of parallel pancake-shaped traps, described by linearly coupled
two-dimensional NPSEs.

This work was partially supported by Fondazione Cariparo (Padova, Italy).
B.A.M. appreciates grant No. 149/2006 from the German-Israel Foundation.

\end{document}